\documentclass[preprint,prc,aps,epsfig]{revtex4}
\usepackage{psfig}

\newcommand{\AmS}{{\protect\the\textfont2
  A\kern-.1667em\lower.5ex\hbox{M}\kern-.125emS}}
\def\beq{\begin{equation}}
\def\eeq{\end{equation}}
\def\beqa{\begin{eqnarray}}
\def\eeqa{\end{eqnarray}}
\def\MeV{\nobreak\,\mbox{MeV}}
\def\GeV{\nobreak\,\mbox{GeV}}

\begin{document}

\title{The $J/\psi-K^*$ dissociation cross section in a meson exchange model}
\author{R.S. Azevedo and M. Nielsen}
\affiliation{Instituto de F\'{\i}sica, 
        Universidade de S\~{a}o Paulo, \\
        C.P. 66318,  05389-970 S\~{a}o Paulo, SP, Brazil}

\begin{abstract}
In this work we study the energy dependence of the $J/\psi-K^*$ 
dissociation cross section, using a meson exchange model based
on effective hadronic Lagrangian that 
includes couplings between pseudoscalar and vector mesons. We also
consider anomalous parity terms.
 Off-shell effects at the vertices were handled 
with QCD sum rule estimates for the running coupling constants, and we
compare the results with and without form factors. We also study the
$J/\psi-\rho$ cross section with form factors.
The total $J/\psi-K^*$ cross section was found to be 
$1.6 \sim1.9$ mb for $4.2\leq\sqrt{s}\leq5~\GeV$.
\end{abstract}

\maketitle

\section{Introduction}

In relativistic  heavy  ion  collisions $J/\psi$ suppression has been
recognized as an important tool to identify the possible phase transition
to quark-gluon  plasma (QGP) \cite{ma86,vo99,ge99}. 
However, even if the QGP is not formed, the $J/\psi$ can be dissociated
by many other ``comoving'' light hadrons which are produced in such 
collisions. Therefore, the evaluation of the size of the $J/\psi$ cross
section by light mesons is important to distinguish wheter the QGP
is formed or the $J/\psi$ is dissociated by other ``comoving'' hadrons.

Since the cross section of $J/\psi$ dissociation by hadrons can not be
directly measured, several theoretical approaches have been proposed to
estimate their values. Some
approaches were based on charm quark-antiquark dipoles interacting with the
gluons of a larger
(hadron target) dipole \cite{bhp,kha2,lo} or quark exchange between two
(hadronic) bags \cite{wongs,mbq}, or QCD sum rules \cite{nnmk,dlnn,dklnn}, 
whereas other works used the meson exchange mechanism
\cite{mamu98,haglin,linko,tho,haga,osl,nnr,haga2,aze}.

The meson exchange approach was applied 
to $J/\psi-N$, $J/\psi-\pi$, $J/\psi-\rho$, and $J/\psi-K$  cross sections 
\cite{mamu98,haglin,linko,tho,haga,osl,nnr,haga2,aze}. In ref.~\cite{aze},  
analyzing the $J/\psi-K$  cross section, we have shown that the 
inclusion of off-shell effects at the vertices, through the introduction of 
QCD sum rule estimates for the form factors, can change the energy dependence
of the absorption cross section in a nontrivial way. A similar result
was obtained in \cite{haga2} for the $J/\psi-\pi$ dissociation cross section.
 In this work we extend the analysis done in ref.~\cite{aze} and
evaluate the $J/\psi-K^*$ cross section using a meson-exchange model,
considering anomalous parity terms as in ref.~\cite{osl,aze} and 
QCD sum rule estimates for the form factors. We also study the
changes in the $J/\psi-\rho$ cross section caused by the inclusion of the 
appropriate form factors.

\section{Effective Lagrangians}

Following refs.~\cite{haglin,linko,haga,osl,aze}, we assume that 
SU(4) symmetry is exact in order to obtain the SU(4) Lagrangian
for the pseudo-scalar and vector mesons. The effective Lagrangians 
relevant to study the dissociation of $J/\psi$ by $K^*$ are:

\begin{eqnarray}
&&\mathcal{L}_{\psi DD}=ig_{\psi
DD}\psi^{\mu} \left (D\partial_{\mu}\bar{D}-(\partial_{\mu}D)\bar{D} \right ),
\label{l1}
\\ \label{psidd}
&&\mathcal{L}_{\psi D^{*}D^{*}}=
ig_{\psi D^{*}D^{*}} \Big [\psi^{\mu} \Big ((\partial_{\mu}D^{*\nu})
\bar{D^{*}_{\nu}}
-D^{*\nu}\partial_{\mu}\bar{D^{*}_{\nu}} \Big )
+ \Big ((\partial_{\mu}\psi^{\nu})D^{*}_{\nu}
\nonumber\\
&&-\psi^{\nu}}\partial_{\mu}D^{*}_{\nu} \Big)\bar{D^{*\mu}}
+D^{*\mu}\Big(\psi^{\nu}\partial_{\mu}\bar{D^{*}_{\nu}}
-(\partial_{\mu}\psi^{\nu})\bar{D^{*}_{\nu} \Big) \Big],
\\
&&\mathcal{L}_{\psi D_{s}D_{s}}=
ig_{\psi D_{s}D_{s}}\psi^{*\mu} \left (D_{s}\partial_{\mu}\bar{D_{s}}
-\partial_{\mu}D_{s}\bar{D_{s}} \right ),
\\
&&\mathcal{L}_{\psi D^{*}_{s}D^{*}_{s}}=
ig_{\psi D^{*}_{s}D^{*}_{s}}\Big[\psi^{\mu} \Big(
(\partial_{\mu}D^{*\nu}_{s})\bar{D^{*}_{s\nu}}
-D^{*\nu}_{s}\partial_{\mu}\bar{D^{*}_{s\nu}}\Big)
+\Big((\partial_{\mu}\psi^{\nu})D^{*}_{s\nu}
\nonumber\\
&&-\psi^{\nu}}\partial_{\mu}D^{*}_{s\nu}\Big)\bar{D^{*\mu}_{s}}
+D^{*\mu}_{s}\Big(\psi^{\nu}\partial_{\mu}\bar{D^{*}_{s\nu}}
-\partial_{\mu}\psi^{\nu}\bar{D^{*}_{s\nu}\Big)\Big],
\\
&&\mathcal{L}_{K^{*} DD_{s}} = ig_{K^{*} DD_{s}}
\left [K^{*}_{\mu} \left (D\partial^{\mu}\bar{D}_{s}-\bar{D}_{s}
\partial^{\mu}D \right )+\bar{K}^{*}_{\mu} \left (D_{s}\partial^{\mu}\bar{D}-
\bar{D}\partial^{\mu}D_{s} \right ) \right ]\\
&&\mathcal{L}_{K^{*} D^{*}D_{s}^{*}}=
ig_{K^{*} D^{*}D_{s}^{*}}\Big[K^{*\mu}\Big((\partial_{\mu}D^{*})
\bar{D}^{*}_{s\nu}
-(\partial_{\mu}\bar{D}_{s}^{*})D^{*}_{\nu}\Big)+\bar{K}^{*\mu}
\Big((\partial_{\mu}D^{*}_{s\nu})\bar{D}^{*}_{\nu}
-(\partial_{\mu}\bar{D}^{*\nu})D^{*}_{s\nu}\Big)\nonumber \\
&&+ D_{s}^{*\mu}\Big((\partial_{\mu}
\bar{D}^{*\nu})\bar{K}^{*}_{\nu}-(\partial_{\mu}\bar{K}^{*\nu})D^{*}_{\nu} 
\Big)
+\bar{D}^{*\mu}_{s}\Big((\partial_{\mu}K^{*\nu})D^{*}_{\nu}-(\partial_{\mu}D^{*\nu})
K^{*\nu}\Big)+ \nonumber \\
&& D^{*\mu}\Big((\partial_{\mu}\bar{D}^{*\nu}_{s})K^{*}_{\nu}
-(\partial_{\mu}K^{*}_{\nu})\bar{D}^{*\nu}_{s}\Big) 
+ \bar{D}^{*\mu}\Big((\partial_{\mu}\bar{K}^{*\nu})D^{*}_{s\nu}
-(\partial_{\mu}D_{s\nu}^{*})\bar{K}^{*\nu}\Big)\Big], \\
&&\mathcal{L}_{K^{*}\psi DD_{s}}=
+g_{K^{*}\psi DD_{s}}\psi^{\mu} \left (K^{*}_{\mu}\bar{D}_{s}D^{*} 
+\bar{K}^{*}_{\mu} D_{s}\bar{D}_{\mu}^{*} \right ),
\\
&&\mathcal{L}_{K^{*}\psi D^{*}D_{s}^{*}}=
-g_{K^{*}\psi D^{*}D_{s}^{*}}\Big[D_{s\mu}^{*}\Big(\psi^{\nu}
\bar{D}^{*}_{\nu}\bar{K}^{*\mu} +\psi^{\mu} \bar{D}^{*\nu}
\bar{K}^{*}_{\nu}-2\psi^{\nu}\bar{D}^{*}_{\mu}\bar{K}^{*}_{\nu}\Big)+ \nonumber \\
&&\bar{D}^{*}_{s\mu}\Big(\psi^{\nu}D^{*}_{\nu}K^{*\mu} +\psi^{\mu}D^{*\nu}
K^{*}_{\nu}-2\psi^{\nu}D^{*\mu}K^{*}_{\nu}\Big)\Big].\label{kpsidds}
\end{eqnarray}
where we have defined the charm meson and $K^*$ iso-doublets 
$D\equiv(D^0,D^+)$,
$D^*\equiv(D^{*0},D^{*+})$ and $K^* \equiv(K^{*0},K^{*+})$. 

The anomalous parity terms, 
introduced in ref.~\cite{osl} for the $J/\psi-\pi$ and 
$J/\psi-\rho$ cases, important for the $J/\psi-K^*$ case are:
\begin{eqnarray}
&& \mathcal{L}_{\psi D^{*}D}=g_{\psi D^{*}D}\epsilon^{\mu \nu \alpha \beta}
\partial_{\mu}\psi_{\nu} \left (\partial_{\alpha}\bar{D}^{*}_{\beta}D
+\partial_{\alpha}D^{*}_{\beta}\bar{D} \right ), \\
&& \mathcal{L}_{\psi D_{s}^{*}D_{s}}=g_{\psi D_{s}^{*}D_{s}}\epsilon^{\mu 
\nu \alpha \beta}\partial_{\mu}\psi_{\nu} \left 
(\partial_{\alpha}\bar{D}^{*}_{s\beta}D_{s}
+\partial_{\alpha}D^{*}_{s\beta}\bar{D}_{s} \right ), \\
&& \mathcal{L}_{K^{*} D_{s}^{*}D}=-g_{K^{*} D_{s}^{*}D}\epsilon^{\mu \nu 
\alpha \beta} \left 
(\partial_{\mu}\bar{K}^{*}_{\nu}\partial_{\alpha}\bar{D}^{*}_{s\beta}D
+\partial_{\mu}\bar{K}^{*}_{\nu}\partial_{\alpha}D^{*}_{s
\beta}\bar{D} \right ),\\
&& \mathcal{L}_{K^{*} D_{s}D^{*}}=-g_{K^{*} D_{s}D^{*}}\epsilon^{\mu \nu 
\alpha \beta} \left 
(\partial_{\mu}\bar{K}^{*}_{\nu}\partial_{\alpha}\bar{D}^{*}_{
\beta}D_{s}+\partial_{\mu}\bar{K}^{*}_{\nu}\partial_{\alpha}D^{*}_{
\beta}\bar{D}_{s} \right),\\
&& \mathcal{L}_{K^{*} D_{s}D^{*}\psi}=ig_{K^{*} D_{s}D^{*}\psi}
\epsilon^{\mu \nu \alpha 
\beta}\psi_{\mu} \left (-\partial_{\beta}\bar{D}_{s}K^{*}_{\alpha}D^{*}_{\nu}
+\partial_{\beta}D_{s}\bar{K}^{*}_{\alpha}\bar{D}^{*}_{\nu} \right )
 \nonumber \\
&&+ih_{K^{*} D_{s}D^{*}\psi}\epsilon^{\mu \nu \alpha 
\beta}\Big[\psi_{\mu}\Big((\partial_{\nu}\bar{K}^{*}_{\alpha})
\bar{D}^{*}_{\beta}D_{s}
-(\partial_{\nu}K^{*}_{\alpha})D^{*}_{\beta}\bar{D}_{s} 
+3(\partial_{\nu}\bar{D}_{\alpha}^{*})\bar{K}^{*}_{\beta}D_{s}
-3(\partial_{\nu}D_{\alpha}^{*})K^{*}_{\beta}\bar{D}_{s}\Big)
 \nonumber \\
&&-\partial_{\nu}\psi_{\mu}\Big(\bar{K}^{*}_{\alpha}\bar{D}^{*}_{\beta}D_{s}
-K^{*}_{\alpha}D^{*}_{\beta}\bar{D}_{s}\Big)\Big] \\
&&\mathcal{L}_{K^{*} D_{s}^{*}D\psi}=ig_{K^{*} D_{s}^{*}D\psi}
\epsilon^{\mu \nu \alpha 
\beta}\psi_{\mu} \left (-\partial_{\beta}\bar{D}\bar{K}^{*}_{\alpha}
D^{*}_{s\nu}
+\partial_{\beta}DK^{*}_{\alpha}\bar{D}^{*}_{s\nu} \right ) \nonumber \\
&&+ih_{K^{*} D_{s}^{*}D\psi}\epsilon^{\mu \nu \alpha 
\beta}\Big[\psi_{\mu}\Big((\partial_{\nu}K^{*}_{\alpha})\bar{D}^{*}_{s\beta}D
-(\partial_{\nu}\bar{K}^{*}_{\alpha})D^{*}_{s\beta}\bar{D} 
-3(\partial_{\nu}D_{s\alpha}^{*})\bar{K}^{*}_{\beta}\bar{D}
+3(\partial_{\nu}\bar{D}_{s\alpha}^{*})K^{*}_{\beta}D\Big) \nonumber \\
&&-\partial_{\nu}\psi_{\mu}\Big(\bar{K}^{*}_{\alpha}D^{*}_{s\beta}\bar{D}
-K^{*}_{\alpha}\bar{D}^{*}_{s\beta}D\Big)\Big],
\end{eqnarray}

\begin{figure}[htb] \label{fig1}
\centerline{\psfig{figure=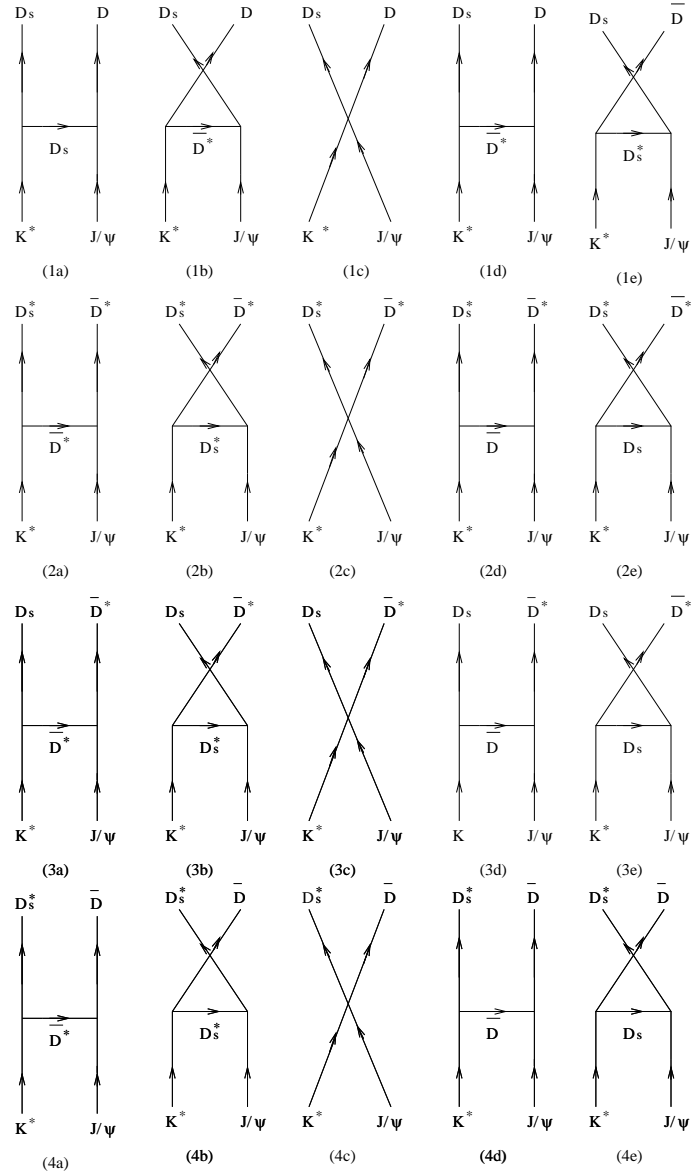,width=9cm,angle=0}}
\vspace{-.5cm}
\caption{\small{Diagrams for $J/\psi$ absorption processes:   
1) $K^* \psi \rightarrow D_s {\bar D}$;
2) $K^* \psi \rightarrow D_s^* {\bar D}^*$;
3) $K^* \psi \rightarrow D_s {\bar D}^*$;
4) $K^* \psi \rightarrow D_s^* \bar D$. Diagrams for
the processes $\bar K^* \psi \rightarrow {\bar D}_s D$,
$\bar K^* \psi \rightarrow {\bar D}_s^* D$, $\bar K^* \psi \rightarrow 
{\bar D}_s D^*$, and
$\bar K^* \psi \rightarrow {\bar D}_s^* D$
are similar to (1a)-(1e) through (4a)-(4e) respectively, but with each 
particle replaced by its anti-particle.}}  
\end{figure}

The processes we are interested in studying for the dissociation of $J/\psi$ 
by $K^*$ are represented by Fig.~1. They  are:
\beqa
K^* J/\psi \rightarrow D_s {\bar D}, && \bar K^* J/\psi \rightarrow D 
{\bar D}_s,
\label{pro1}\\ 
K^* J/\psi \rightarrow D_s^* {\bar D}^*, && \bar K^* J/\psi \rightarrow D^* 
{\bar D}_s^*,
\label{pro2}\\
K^* J/\psi \rightarrow D_s {\bar D}^*, && \bar K^* J/\psi \rightarrow D^* 
{\bar D}_s.
\label{pro3}\\
K^* J/\psi \rightarrow D_s^* {\bar D}, && \bar K^* J/\psi \rightarrow D 
{\bar D}_s^*,
\label{pro4}
\eeqa
Since the two processes in eqs.~(\ref{pro1}), (\ref{pro2}), (\ref{pro3}) and 
(\ref{pro4}) have the same 
cross section, in Fig.~1 we only show the diagrams for the first process
in eqs.~(\ref{pro1}) through (\ref{pro4}).

Defining  the four-momentum of the $K^*$ and 
the $J/\psi$  by $p_1$, $p_2$ respectively, and the four-momentum of the
final-state mesons by $p_3$ and $p_4$, the full amplitude for the 
processes $K^* \psi \rightarrow D_s {\bar D}$, shown in diagrams (1) of 
Fig.~1, 
without isospin factors and before summing and averaging over external 
spins, is given by

\begin{eqnarray}
{\cal M}_1 \equiv {\cal M}_1^{\mu \nu} 
~\epsilon_{1 \mu} \epsilon_{2 \nu}^* 
=\left ( \sum_{j=a,b,c,d,e} {\cal M}_{1j}^{\mu \nu} \right )
\epsilon_{1 \mu} \epsilon_{2 \nu},  
\label{m1}
\end{eqnarray}
with
\begin{eqnarray}
&& M^{\mu \nu}_{1a} = -g_{\psi DD}g_{K^{*}D_{s}D} \left (p_{1} - 2p_{3} 
\right )^{\mu} \left (\frac{1}{t - m_{D}^{2}} \right ) \left (p_{4} - p_{3} 
+ p_{1} \right )^{\nu} , \nonumber \\
&& M^{\mu \nu}_{1b} = -g_{K^{*}D_{s}D}g_{\psi
D_{s}D_{s}}\left (-p_{1} + 2p_{4} \right )^{\mu} \left (\frac{1}{u 
- m_{\bar{D}_{s}}^{2}} \right ) \left (-p_{1} 
- p_{3} + p_{4} \right )^{\nu}, \nonumber \\
&& M^{\mu \nu}_{1c} = g_{K^{*}\psi DD_{s}}g^{\mu \nu}, \nonumber  \\
&& M^{\mu \nu}_{1d} = -\frac{g_{K^{*}D_{s}D^{*}}
g_{\psi D^{*}D}}{t-m_{\bar{D}^{*}}^{2}}\epsilon^{\mu  
\sigma \rho \alpha}\epsilon^{\nu \delta \chi \beta} \left (g_{\alpha \beta} 
- \frac{(p_{1} - p_{3})_{\alpha}(p_{1} 
- p_{3})_{\beta}}{m_{\bar{D}^{*}}^{2}} \right )p_{1\sigma}p_{2\delta}p_{3
\rho}p_{4\chi}, \nonumber \\
&&M^{\mu \nu}_{1e} = -\frac{g_{K^{*}D_{s}^{*}D}
g_{\psi D^{*}_{s}D_{s}}}{u-m_{\bar{D}^{*}_{s}}^{2}}\epsilon^{\mu  
 \rho \sigma \alpha}\epsilon^{\nu \delta \chi \beta} \left (g_{\alpha \beta} 
- \frac{(p_{2} - p_{3})_{\alpha}(p_{2} 
- p_{3})_{\beta}}{m_{\bar{D}^{*}_{s}}^{2}} \right )p_{1\sigma}p_{2\delta}p_{3
\chi}p_{4\rho},\label{m1e}
\end{eqnarray}
where $t=(p_1-p_3)^2$ and $u=(p_1-p_4)^2$.

Similarly,  the full amplitude for the 
processes $K^* \psi \rightarrow D_s^* {\bar D}^*$, shown in diagrams (2) of 
Fig.~1, 
without isospin factors and before summing and averaging over external 
spins, is given by
\begin{eqnarray}
{\cal M}_2 \equiv {\cal M}_2^{\mu \nu \lambda \gamma } 
~\epsilon_{1 \mu} \epsilon_{2 \nu} \epsilon_{3\lambda}^{*} 
\epsilon_{4\gamma}^{*}
=\left ( \sum_{j=a,b,c,d,e} {\cal M}_{2j}^{\mu \nu \lambda \gamma} \right )
\epsilon_{1 \mu} \epsilon_{2 \nu} \epsilon_{3\lambda}^{*} 
\epsilon_{4\gamma}^{*},  
\label{m2}
\end{eqnarray}
where
\begin{eqnarray}
&& M^{\mu \nu \lambda \gamma}_{2a} = g_{\psi D^{*}D^{*}}
g_{K^{*}D^{*}D_{s}^{*}} \left [2p_{3}^{\mu}g^{\alpha \lambda} 
+ (-p_{1} - p_{3})^{\alpha}g^{\mu \lambda} + 2p_{1}^{
\lambda}g_{\alpha \mu} \right ] \left (\frac{1}{t - m_{D^{*}}^{2}} \right )
 \nonumber \\
&&\left (g_{\alpha \beta} - \frac{(p_{1} - p_{3})_{\alpha}(p_{1} - p_{3})_{\beta}}{
m_{D^{*}}^{2}} \right ) \left [-2p_{2}^{\gamma}g^{\beta \nu} 
+ (p_{2} + p_{4})^{\beta}g^{\nu \gamma} - 2p_{4}^{
\nu}g_{\gamma \beta} \right ] \nonumber \\
&& M^{\mu \nu \lambda \gamma}_{2b} = g_{K^{*}D^{*}D_{s}^{*}}g_{\psi
D^{*}_{s}D^{*}_{s}} \left [-2p_{1}^{\gamma}g^{\alpha \mu} 
+ (p_{1} + p_{4})^{\alpha}g^{\gamma \mu} + 2p_{4}^{
\mu}g_{\alpha \gamma} \right ] \left (\frac{1}{u - m_{D_{s}^{*}}^{2}} \right )
 \nonumber \\
&& \left (g_{\alpha \beta} - \frac{(p_{1} - p_{4})_{\alpha}(p_{1} - p_{4})_{\beta}}{
m_{D_{s}^{*}}^{2}} \right ) \left [2p_{2}^{\lambda}g^{\nu \beta} 
- (p_{2} + p_{3})^{\beta}g^{\nu \lambda} + 2p_{3}^{
\nu}g_{\beta \lambda} \right ], \nonumber \\
&& M^{\mu \nu \lambda \gamma}_{2c} = g_{K^{*}\psi D^{*}D_{s}^{*}}
\left (g^{\nu \lambda}g^{\mu \gamma} + g^{\nu \gamma}g^{\mu \lambda}
-2g^{\nu \mu}g^{\gamma \lambda} \right ), \nonumber \\
&& M^{\mu \nu \lambda \gamma}_{2d} = -\frac{g_{K^{*}D_{s}^{*}D}
g_{\psi D^{*}D}}{t-m_{D}^{2}}\epsilon^{\mu \lambda \rho \sigma}
\epsilon^{\nu \gamma 
\delta \chi}p_{1\sigma}p_{2\delta}p_{3
\rho}p_{4\chi}, \nonumber \\
&&M^{\mu \nu \lambda \gamma}_{2e} = -\frac{g_{K^{*}D_{s}D^{*}}
g_{\psi D_{s}^{*}D_{s}}}{u-m_{D_{s}}^{2}}
\epsilon^{\mu \gamma \sigma \rho}\epsilon^{\nu \lambda 
\delta \chi}p_{1\sigma}p_{2\delta}p_{3s
\chi}p_{4\rho}.
\end{eqnarray}

Calling the four-momentum of the pseudo-scalar  and the vector final-state 
mesons  by $p_3$ and $p_4$ respectively, the full amplitude for the 
processes $K^* \psi \rightarrow D_s {\bar D}^*$ shown in diagram (3) 
of Fig.~1 is 
\begin{equation}
{\cal M}_{3} \equiv {\cal M}_{3}^{\mu \nu \gamma}\epsilon_{1\mu}
\epsilon_{2\nu}\epsilon_{4\gamma}^{*}=
\left(\sum_{i=a,b,c,d,e}{\cal M}_{3i}^{\mu \nu \gamma}\right)
\epsilon_{1\mu}\epsilon_{2\nu}\epsilon_{4\gamma}^{*},
\end{equation}
with
\begin{eqnarray}
&& M^{\mu \nu \gamma }_{3a} = -\frac{g_{K^{*}D_{s}D^{*}}g_{\psi D^{*}D^{*}}}{t
-m_{\bar{D}^{*}}^{2}}\epsilon^{\mu \rho \sigma \alpha} \left (g_{\alpha \beta} 
- \frac{(p_{1} - p_{3})_{\alpha}(p_{1} - p_{3})_{
\beta}}{m_{\bar{D}^{*}}^{2}} \right )p_{1\sigma}p_{3\rho} \nonumber \\
&& \left [2p_{2}^{\gamma}g^{\beta \nu} 
+ (-p_{2} - p_{4})^{\beta}g^{\gamma \nu} + 2p_{4}^{
\nu}g_{\beta \gamma} \right ], \nonumber \\
&&  M^{\mu \nu \gamma }_{3b} = -\frac{g_{K^{*}D_{s}^{*}D^{*}}
g_{\psi D_{s}^{*}D_{s}}}{u-m_{D_{s}^{*}}^{2}}
\epsilon^{\nu \delta \chi \beta}\left (g_{\alpha \beta} 
- \frac{(p_{1} - p_{4})_{\alpha}(p_{1} - p_{4})_{
\beta}}{m_{D_{s}^{*}}^{2}} \right )p_{2\delta}p_{3\chi} \nonumber \\
&& \left [2p_{1}^{\gamma}g^{\alpha \mu} 
+ (-p_{1} - p_{4})^{\alpha}g^{\gamma \mu} + 2p_{4}^{
\mu}g_{\alpha \gamma} \right ], \nonumber  \\
&& M^{\mu \nu \gamma }_{3c} = g_{K^{*}D^{*}D_{s}\psi}\epsilon^{\mu \nu \gamma 
\delta}p_{3\delta}+h_{K^{*}D^{*}D_{s}\psi}\epsilon^{\mu \nu \gamma \delta}
(p_{1}+3p_{4}-p_{2})_{\delta}, \nonumber \\
&&M^{\mu \nu \gamma }_{3d}=-2\frac{g_{K^{*}D_{s}D}
g_{\psi D^{*}D}}{t-m_{D}^{2}}\epsilon^{\nu \gamma \delta \chi}
p_{2\delta}p_{4\chi}p_{3}^{\mu}, \nonumber \\
&&M^{\mu \nu \gamma }_{3e}=-2\frac{g_{K^{*}D_{s}D^{*}}
g_{\psi D_{s}D_{s}}}{u-m_{D_{s}}^{2}}\epsilon^{\mu \gamma \rho \sigma}
p_{1\sigma}p_{4\rho}p_{3}^{\nu}.
\end{eqnarray}

For the diagram (4) in Fig.~1, representing the processes 
$K^* \psi \rightarrow D_s^* {\bar D}$, calling the four-momentum of the 
 vector and pseudoscalar final-state mesons
respectively by $p_3$ and $p_4$, the full amplitude is given by
\begin{equation}
{\cal M}_{4} \equiv {\cal M}_{4}^{\nu \lambda \mu}\epsilon_{1\mu} 
\epsilon_{2\nu}\epsilon_{3\lambda}^{*}=\left(\sum_{i=a,b,c,d,e}
{\cal M}_{4i}^{\nu \lambda \mu}\right)
\epsilon_{1\mu} \epsilon_{2\nu} \epsilon_{3\lambda}^{*}, 
\end{equation}
with
\begin{eqnarray}
&& M_{4a}^{\mu \nu \lambda} =-\frac{g_{K^{*}D_{s}^{*}D^{*}}g_{\psi D^{*}D}}{t
-m_{\bar{D}^{*}}^{2}}\epsilon^{\nu \chi \delta \beta} \left (g_{\alpha \beta} 
- \frac{(p_{1} - p_{3})_{\alpha}(p_{1} - p_{3})_{
\beta}}{m_{\bar{D}^{*}}^{2}} \right )p_{2\delta}p_{4\chi} \nonumber \\
&& \left [2p_{1}^{\lambda}g^{\alpha \mu} 
+ (-p_{1} - p_{3})^{\alpha}g^{\lambda \mu} + 2p_{3}^{
\mu}g_{\alpha \lambda} \right ], \nonumber \\
&& M_{4b}^{\mu \nu \lambda}=\frac{g_{K^{*}DD_{s}^{*}}g_{\psi
D_{s}^{*}D_{s}^{*}}}{u - m_{D_{s}^{*}}^{2}} \left (g^{\alpha \beta} 
- \frac{(p_{1} - p_{4})^{\alpha}(p_{1} - p_{4})^{\beta}}{m_{D_{s}^{*}}^{2}} 
\right )
\epsilon_{\mu \sigma \rho \alpha}p_{1\sigma}p_{4\rho} \nonumber \\
&& \left [2p_{2}^{\lambda}g^{\beta \nu} 
+ (-p_{2} - p_{3})^{\beta}g^{\lambda \nu} + 2p_{3}^{
\nu}g_{\beta \lambda} \right ], \nonumber \\
&&  M_{4c}^{\mu \nu \lambda }=-g_{K^{*}DD_{s}^{*}\psi}
\epsilon^{\mu \nu \lambda \delta}p_{4\delta}
-h_{K^{*}D^{*}D_{s}\psi}\epsilon^{\mu \nu \lambda \delta}
(p_{1}+3p_{3}-p_{2})_{\delta}, \nonumber \\
&&M^{\mu \nu \lambda }_{4d}=-2\frac{g_{K^{*}D_{s}^{*}D}
g_{\psi DD}}{t-m_{D}^{2}}\epsilon^{\mu \lambda \sigma \rho}
p_{1\sigma}p_{4\rho}p_{4}^{\nu}, \nonumber \\
&&M^{\mu \nu \lambda }_{4e}=-2\frac{g_{K^{*}D_{s}D}
g_{\psi D_{s}^{*}D_{s}}}{u-m_{D_{s}}^{2}}\epsilon^{\nu \lambda \chi \delta}
p_{2\delta}p_{3\chi}p_{4}^{\mu}. \label{m4e}
\end{eqnarray}

After averaging (summing) over initial (final) spins 
and including isospin factors, the cross sections 
for these four processes are given by 
\begin{eqnarray}
&&\frac {d\sigma_1}{dt}= \frac {1}{576 \pi s p_{i,\rm cm}^2} 
 M_1^{\mu \nu}M_1^{*\mu^\prime \nu^\prime}
\left ( g_{\mu \mu^\prime}-\frac{p_{1\mu} p_{1 \mu^\prime}} {m_1^2} \right )
\left ( g_{\nu \nu^\prime}
-\frac{p_{2 \nu} p_{2 \nu^\prime}} {m_2^2} \right ), \label{jkaon12} \\
&&\frac {d\sigma_2}{dt}= \frac {1}{576 \pi s p_{i,\rm cm}^2} 
 M_2^{\mu \nu \lambda \gamma}M_2^{*\mu^\prime \nu^\prime \lambda^\prime 
\gamma^\prime}
\left ( g_{\mu \mu^\prime}-\frac{p_{1\mu} p_{1 \mu^\prime}} {m_1^2} \right )
\left ( g_{\nu \nu^\prime}
-\frac{p_{2 \nu} p_{2 \nu^\prime}} {m_2^2} \right )
\nonumber \\
&&\times
\left ( g_{\lambda \lambda^\prime}
-\frac{p_{3 \lambda} p_{3 \lambda^\prime}} {m_3^2} \right )\left ( g_{\gamma 
\gamma^\prime}
-\frac{p_{4\gamma} p_{4\gamma^\prime}} {m_4^2} \right ), \label{jkaon2} \\
&&\frac {d\sigma_3}{dt}= \frac {1}{576 \pi s p_{i,\rm cm}^2} 
 M_3^{\mu \nu \gamma} M_3^{*\mu^\prime \nu^\prime \gamma^\prime}
\left ( g_{\mu \mu^\prime}-\frac{p_{1\mu} p_{1\mu^\prime}} {m_1^2} \right )
\left ( g_{\nu \nu^\prime}-\frac{p_{2 \nu} p_{2 \nu^\prime}} {m_2^2} \right )
\nonumber \\
&&\times
\left ( g_{\gamma \gamma^\prime}
-\frac{p_{4\gamma} p_{4\gamma^\prime}} {m_4^2} \right ),\label{jkaon3}
\end{eqnarray}and
\begin{eqnarray}\label{jkaon14}
&&\frac {d\sigma_4}{dt}= \frac {1}{576 \pi s p_{i,\rm cm}^2} 
  M_4^{\mu \nu \lambda} M_4^{*\mu^\prime \nu^\prime \lambda^\prime}
\left ( g_{\mu \mu^\prime}-\frac{p_{1\mu} p_{1\mu^\prime}} {m_1^2} \right )
\left ( g_{\nu \nu^\prime}-\frac{p_{2 \nu} p_{2 \nu^\prime}} {m_2^2} \right )
\nonumber\\
&&\times
\left ( g_{\lambda \lambda^\prime}
-\frac{p_{3\lambda} p_{3\lambda^\prime}} {m_3^2} \right ),
\end{eqnarray}
with $s=(p_1+p_2)^2$, and  
\begin{eqnarray}
p_{0,\rm cm}^2=\frac {\left [ s-(m_1+m_2)^2 \right ]
\left [ s-(m_1-m_2)^2 \right ]}{4s}
\end{eqnarray}
being the three-momentum squared of initial-state mesons in the 
center-of-momentum (c.m.) frame. 

\section{Results}

Exact SU(4) symmetry 
would give the following relations among the coupling constants 
\cite{linko,osl,aze}:
\begin{eqnarray}
&&g_{K^* D_sD}=g_{K^* D^*D_s^*}=\frac{g}{2\sqrt{2}}~, \nonumber \\
&&g_{\psi DD}=g_{\psi D_sD_s}=g_{\psi D^* D^*}=g_{\psi D_s^* D_s^*}=
\frac{g}{\sqrt 6}~, \nonumber \\
&&g_{\psi K^* D_sD}=\frac{g^2}{2 \sqrt 3}~,\nonumber \\
&&g_{\psi K^* D_s^*D^*}=\frac{g^2}{4 \sqrt 3}~,\nonumber \\
&& g_{\psi D^{*}D}=g_{\psi D_{s}^{*}D_{s}}=
{\sqrt{2}g_a^2N_c\over 64\sqrt{3}\pi^2F_\pi}
, \nonumber \\
&&g_{K^*DD_{s}^*}=g_{K^*D^*D_{s}}={\sqrt{2}g_a^2N_c \over 64\pi^2F_\pi}, 
\nonumber \\
&&g_{\psi K^*D^*D_{s}}=g_{\psi K^*DD_{s}^{*}}=
3~h_{\psi K^*D^*D_{s}}=3~h_{\psi K^*DD_{s}^{*} }=\frac{\sqrt{3}g_a^3N_c}
{256\pi^2F_\pi^3}, 
\label{su4}
\end{eqnarray}
where $N_c=3$ and $F_\pi=132\MeV$.

None of the above couplings are known experimentally, and one has to use
models to estimate them.  In refs.~\cite{haga2,aze}, 
the $J/\psi-\pi$, $J/\psi-\rho$ and  $J/\psi- k$ cross sections
were evaluated by using form factors and coupling constants estimated 
using QCD sum rules \cite{nos1,nos2,nos3,nos4}. The results in 
ref.~\cite{haga2,aze} show that,
with  appropriate form factors, even the behavior of the cross section
as a function of $\sqrt{s}$ can change. In this work
we use the form factors in the vertices ${J/\psi DD}$, ${J/\psi D^{*}D}$
and $\rho DD$, determined from QCD sum rules \cite{nos2,nos5}, and the 
above SU(4) relations to 
estimate the form factors and coupling constants in all vertices.

From ref.~\cite{nos5} we get $g_{\psi DD^{*}}=4.0~\GeV^{-1}$ and
$g_{\psi DD}=5.8$. Using these numbers in the SU(4) relations given in
Eqs.~(\ref{su4}) we obtain
\begin{eqnarray}
&&g_{\psi DD}=g_{\psi D_sD_s}=g_{\psi D^* D^*}=g_{\psi D_s^* D_s^*}=
5.8,\;\;\;
g_{K^* D_sD}=g_{K^* D^*D_s^*}=5.0,\nonumber \\
&&g_{\psi K^* D_sD}=
58.2,\;\;\;g_{\psi K^* D_s^*D^*}=29.0,\;\;\;g_{\psi D^{*}D}=
g_{\psi D_{s}^{*}D_{s}}=4.0~\hbox{GeV}^{-1}, \nonumber \\
&& g_{K^*D_{s}^{*}D}=g_{K^*D_{s}D^*}=6.9~\hbox{GeV}^{-1}, \nonumber \\
&&g_{\psi K^*D^*D_{s}}=g_{\psi K^*DD_{s}^*}=24.8~\hbox{GeV}^{-1},\nonumber \\
&& h_{\psi K^* D^{*}D_{s}}= h_{\psi K^* DD_{s}^*}=8.3~\hbox{GeV}^{-1}.
\label{sun}
\end{eqnarray}

The form factors given in ref.~\cite{nos5} are
\beq
g^{(D^{*})}_{\psi DD^{*}}(t)=g_{\psi DD^{*}}
\left(5~e^{-\left(\frac{27-t}{18.6}\right)^2}\right)=g_{\psi DD^{*}}~h_1(t),
\label{h1}
\eeq
\beq
g^{(D)}_{\psi DD^{*}}(t)=g_{\psi DD^{*}}\left(3.3
~e^{-\left(\frac{26-t}{21.2}\right)^2}\right)=g_{\psi DD^{*}}~h_2(t),
\label{h2}
\eeq
\beq
g^{(D)}_{\psi DD}(t)=g_{\psi DD}
\left(2.6~e^{-\left(\frac{20-t}{15.8}\right)^2}\right)=g_{\psi DD}~h_3(t),
\label{h3}
\eeq
where $g_{123}^{(1)}$ means the form factor at the vertex involving the
mesons $123$ with the meson $1$ off-shell. In the above equations the
numbers in the exponentials are in units of $\GeV^2$.
Since there is no QCD sum rule calculation for the form factors at the
vertices ${K^*D_s D}$, we make the supposition that they are
similar to the form factor at the vertex $\rho DD$. From ref.~\cite{nos4}
we get
\begin{equation}
g^{(D)}_{\rho DD}(t)=g_{\rho DD}\left({(3.5~\hbox{GeV})^2-m_D^2\over 
(3.5~\hbox{GeV})^2-t}
\right)=g_{\rho DD}~h_4(t,m_D^2).
\label{h4}
\end{equation}
With these form factors the amplitudes will be modified in the following
way: 
\beq
{\cal M}_{ia}\rightarrow h_3(t)h_4(t,m_{ia}^2){\cal M}_{ia},\;\;\;
{\cal M}_{ib}\rightarrow h_3(u)h_4(u,m_{ib}^2){\cal M}_{ib},
\label{ff1}
\eeq
for $i=1,2$, with  $m_{1a}=m_{D},~m_{1b}=m_{D_s},~
m_{2a}=m_D^*,$ and $m_{2b}=m_{D_s^*}$.

\begin{eqnarray}
&&M_{ic}\rightarrow {1\over2}\left(h_3(t)h_4(t,m_{ia}^2)+h_3(u)
h_4(u,m_{ib}^2)\right)M_{ic},\;\;\;
M_{1d}\rightarrow h_1(t)h_4(t,m_{1d}^2)M_{1d}, \nonumber \\
&&M_{ie}\rightarrow h_1(u)h_4(u,m_{ie}^2)M_{ie},\;\;\;M_{2d}\rightarrow 
h_2(t)h_4(t,m_{1e}^2)M_{2d}, \nonumber \\
&&M_{2e}\rightarrow h_2(u)h_4(u,m_{1e}^2)M_{2e},
\end{eqnarray}
for $i=1,2$ with $m_{1d}=m_{D^*}$,$m_{1e}=m_{D_s^*}$, $m_{2d}=m_{D}$ and 
$m_{2e}=m_{D_s}$.
\begin{eqnarray}
&&M_{3a}\rightarrow h_3(t)h_4(t,m_{D^*}^2)M_{3a},\;\;\;
M_{3b}\rightarrow h_1(u)h_4(u,m_{D_s^*}^2)M_{3b}\nonumber\\
&&M_{3c}\rightarrow {1\over4}\left(h_3(t)h_4(t,m_{D^*}^2)+
h_1(u)h_4(u,m_{D_s^*}^2) + h_2(t)h_4(t,m_{D}^2))+ 
h_3(u)h_4(u,m_{D_s}^2)\right),\nonumber \\
&&M_{3d}\rightarrow h_2(t)h_4(t,m_{D}^2)M_{3d},\;\;\;
M_{3e}\rightarrow h_3(u)h_4(u,m_{D_s}^2)M_{3e}
\end{eqnarray}
and
\begin{eqnarray}
&&M_{4a}\rightarrow h_1(t)h_4(t,m_{D^*}^2)M_{4a},\;\;\;
M_{4b}\rightarrow h_3(u)h_4(u,m_{D_s^*}^2)M_{4b},\nonumber \\
&&M_{4c}\rightarrow{1\over4}\left(h_1(t)h_4(t,m_{D^*}^2)
+h_3(u)h_4(u,m_{D_s^*}^2)+h_3(t)h_4(t,m_{D}^2)+h_1(u)
h_4(u,m_{D_s}^2)\right)M_{4c} \nonumber \\
&&M_{4d}\rightarrow h_3(t)h_4(t,m_{D}^2)M_{4d},\;\;\;
M_{4e}\rightarrow h_1(u)h_4(u,m_{D_s}^2)M_{4e}.\label{ff2}
\end{eqnarray}

One can argue that our prescription to introduce the form factors in Eqs.~
(\ref{ff1}) through (\ref{ff2}) might spoil the current conservation associated
with the $J/\psi$ current. However, this is not the case. Let us consider, for
instance, the full amplitude associated with the processes represented by 
diagrams (1) in Fig. 1: ${\cal M}_1^{\mu \nu}= 
= \sum_{j=a,b,c,d,e} {\cal M}_{1j}^{\mu \nu}$. Keeping only terms that will
contribute to the cross section, it can be written as 
\begin{eqnarray}
&&{\cal M}_1^{\mu \nu}=\Lambda_{1}p_{2}^{\mu}p_{1}^{\nu}+
\Lambda_{2}p_{2}^{\mu}p_{3}^{\nu}+\Lambda_{3}p_{4}^{\mu}p_{1}^{\nu}
+\Lambda_{4}p_{4}^{\mu}p_{3}^{\nu}+\Lambda_{5}p_{3}^{\mu}p_{4}^{\nu} 
\nonumber \\
&&+\Lambda_{6}p_{3}^{\mu}p_{1}^{\nu}+
+\Lambda_{7}p_{2}^{\mu}p_{4}^{\nu}+
\Lambda_{8}g^{\mu \nu}, \label{mmodified}
\end{eqnarray}where, before introducing the form factors, these $\Lambda$'s
depend only on the coupling constants and masses.

Without interfering in the final result for the cross section, the amplitude
in Eq. (\ref{mmodified}) can be rewritten as

\begin{eqnarray}
&&{\cal M}_1^{\mu \nu}=\Lambda_{1}\left (p_{1}^{\nu} - \frac{p2.p1}{m^2_{J/\psi}}p_{2}^{\nu} 
\right )p_{2}^{\mu}+
\Lambda_{2}\left (p_{3}^{\nu} - \frac{p2.p3}{m^2_{J/\psi}}p_{2}^{\nu} 
\right )p_{2}^{\mu}+\Lambda_{3}\left (p_{1}^{\nu} - \frac{p2.p1}{m^2_{J/\psi}}p_{2}^{\nu} 
\right )p_{4}^{\mu} \nonumber \\
&&+\Lambda_{4}\left (p_{3}^{\nu} - \frac{p2.p3}{m^2_{J/\psi}}p_{2}^{\nu} 
\right )p_{4}^{\mu}+\Lambda_{5}\left (p_{4}^{\nu} - \frac{p2.p4}{m^2_{J/\psi}}p_{2}^{\nu} 
\right )p_{3}^{\mu} 
+\Lambda_{6}\left (p_{1}^{\nu} - \frac{p2.p1}{m^2_{J/\psi}}p_{2}^{\nu} 
\right )p_{3}^{\mu} \nonumber \\
&&+\Lambda_{7}\left (p_{4}^{\nu} - \frac{p2.p4}{m^2_{J/\psi}}p_{2}^{\nu} 
\right )p_{2}^{\mu}+
\Lambda_{8} \left ( g^{\mu \nu} - \frac{p2^\mu p2^\nu}{m^2_{J/\psi}} 
\right ), \label{mmodified2}
\end{eqnarray}which is explicitly gauge invariant independently of the values
of the parameter $\Lambda_{i}$. Therefore, our prescription in keeping gauge
invariance when the form factors from Eqs (\ref{ff1}) to (\ref{ff2}) are
introduced, is to introduce new terms, proporcional to $p_{2}^{\nu}$, in the 
amplitude, as in Eq. (\ref{mmodified2}). A different prescription can be 
found in ref. \cite{haga2}.

We first analyze the  $J/\psi$ dissociation cross sections by $K^*$ without
considering the form factors, {\it i.e.}, we use the expressions for the 
amplitudes in Eqs.~(\ref{m1e}) through (\ref{m4e}). We will be always
including the contributions for both $J/\psi K^*$ and $J/\psi\bar{K}^*$.
\begin{figure}[htb]
\centerline{\psfig{figure=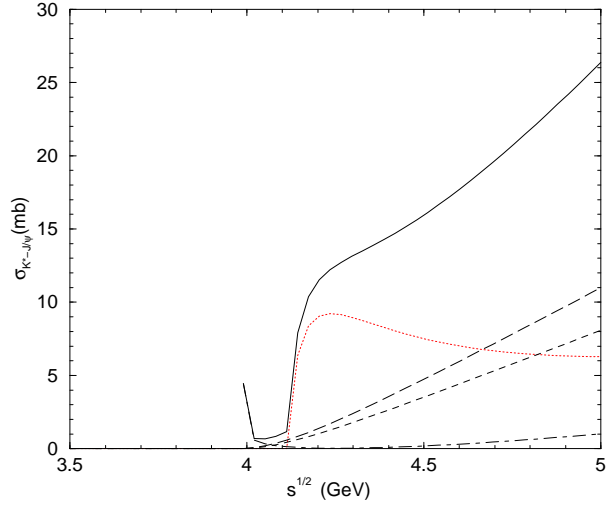,width=8cm}}
\vspace{-.5cm}
\caption{\small{Total cross sections, without form factors,
 for the processes $J/\psi~K^*+J/\psi~\bar{K}^*
\rightarrow$ $\bar{D}~D_s + D~\bar{D}_s$ (dot-dashed line), $\bar{D}^*~
D_s^* +\bar{D}_s^*~D^*$ (dotted line), 
$\bar{D}~D_s^* + D~\bar{D}_s^*$ (dashed line)
and $\bar{D}^*~D_s + D^*~\bar{D}_s$ (long-dashed line).
The solid line gives the total $J/\psi$ dissociation by $K^*$ cross 
section.}}
\label{fig2}
\end{figure}

In Fig. 2 we show the cross section of $J/\psi$ dissociation by $K^*$
as a function of the initial  energy $\sqrt{s}$. 
The dot-dashed, dotted, dashed, long-dashed and solid lines give the
contributions for the processes $J/\psi~K^*+J/\psi~\bar{K}^*
\rightarrow$ $\bar{D}~D_s + D~\bar{D}_s$, $\bar{D}^*~
D_s^* +\bar{D}_s^*~D^*$, $\bar{D}~D_s^* + D~\bar{D}_s^*$,
$\bar{D}^*~D_s + D^*~\bar{D}_s$ and total respectively. 
From this figure we can see that the process $J/\psi~K^*+J/\psi~\bar{K}^*
\rightarrow ~ \bar{D}~D_s + D~\bar{D}_s$ 
has a different dependence with 
the energy near the threshold as compared with the processes
$J/\psi~K^*+J/\psi~\bar{K}^*
\rightarrow ~ \bar{D}~D_s^* + D~\bar{D}_s^*+ \bar{D}^*~D_s + D^*~\bar{D}_s$ 
and $J/\psi~K^*+J/\psi~\bar{K}^*
\rightarrow ~ \bar{D}^*~D_s^* +\bar{D}_s^*~D^*$.
This difference
in the behaviour of the cross section can be understood by noticing 
that while for the process
$J/\psi~K^*+J/\psi~\bar{K}^*~\rightarrow ~ \bar{D}~D_s +\bar{D}_s~D$, 
$m_\psi+m_{K^*}-(m_D+m_{D_s})$ is positive, for the other processes
the mass difference between the initial and final states
is negative. This means that the process 
$J/\psi~K^*+J/\psi~\bar{K}^*~\rightarrow ~\bar{D}~D_s + D~\bar{D}_s$ 
is exotermic and can happen even with $K^*$ and $J/\psi$ at rest. The 
other  processes are endotermic, {\it i.e.},
the reaction only occurs if either $J/\psi$ or $K^*$ have some 
initial energy.

We see also that until $\sqrt{s}\sim4.6~\GeV$ the process 
$J/\psi~K^*+J/\psi~\bar{K}^*~
\rightarrow$ $\bar{D}^*~D_s^* + D^*~\bar{D}_s^*$ dominates. However,
for higher values of $\sqrt{s}$ the processes given by diagrams (3) and (4) 
in Fig.~1 are the most important ones. This is similar to what was
found in ref.~\cite{osl} for the $J/\psi$ dissociation by $\rho$ mesons.

\begin{figure}[htb]
\centerline{\psfig{figure=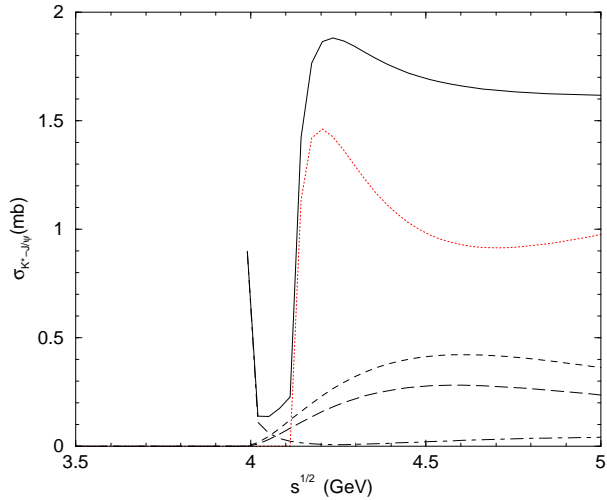,width=8cm}}
\vspace{-.5cm}
\protect\caption{\small{Same as in Fig.~2 but with form factors.}}
\label{fig3}
\end{figure}
In Fig.~3 we show the same processes considered in Fig.~2, but with
form factors. This means that we are now using the amplitudes given by 
Eqs.~(\ref{ff1}) through (\ref{ff2}). The first important conclusion
is that the use of appropriate form factors do change the behavior
of the cross section as a function of $\sqrt{s}$, as obtained in refs.
\cite{haga2,aze}.
The processes more affected by this change are the ones
represented by diagrams (3) and (4) in Fig.~1.
While the total cross section obtained without form factors show a very strong
grown with $\sqrt{s}$, this is not more the case when the total
cross section is obtained with form factors, as can be seen in Fig.~4,
where we show  the total cross section evaluated with and without form factors.
\begin{figure}[htb]
\centerline{\psfig{figure=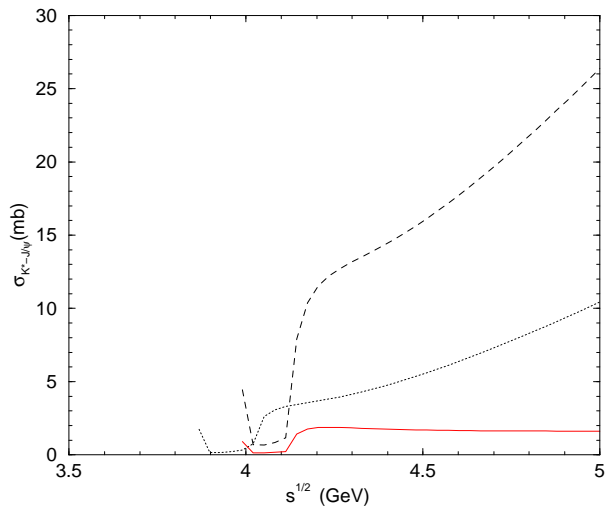,width=8cm}}
\vspace{-.5cm}
\protect\caption{\small{Total $J/\psi$  dissociation cross section 
as a function of the initial energy. The solid and dashed lines give the
results for  $J/\psi$  absorption by $K^*$
with and without form factors respectively. The dotted line gives the results
for  $J/\psi$  absorption by $\rho$ with form factors.}}
\label{fig4}
\end{figure}

One important result of our calculation is the fact that, using
appropriate form factors with cut-offs of order of $\sim3.5~\GeV$ 
(see Eqs.(\ref{h1})
through (\ref{h4})), the value of the cross section  can be reduced
by one order of magnitude. The same effect was obtained in 
refs.~\cite{linko,osl} using monopole form factors, but with cut-offs of 
order of $\sim1~\GeV$, which are considered very small for charmed mesons.

In Fig.~4 we also show, for comparison, 
the total cross section for $J/\psi$  absorption by 
$\rho$'s (dotted line)
using the same form factors and coupling constants given here, and
the value $4.3$ for the $\rho DD$ coupling constant obtained using
QCD sum rule \cite{nos4}.

Other important result that we can see by this figure is that, even with 
form factors, the $J/\psi~\rho$ dissociation cross section still 
increases with $\sqrt{s}$, differently to what happens with the
$J/\psi~K^*$ dissociation cross section. The reason for that is the fact that,
without form factors
the  $J/\psi~\rho$ dissociation cross section increases, with  $\sqrt{s}$,
in a much stronger way than the $J/\psi~K^*$ dissociation cross section,
as can be seen by Fig.~5. 

\begin{figure}[htb]
\centerline{\psfig{figure=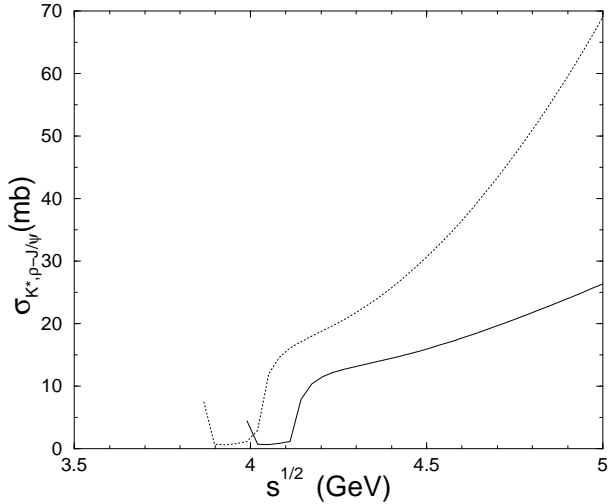,width=8cm}}
\vspace{-.5cm}
\caption{\small{Comparison between the  $J/\psi~K^*$ dissociation cross 
section (solid line) and  $J/\psi~\rho$ dissociation cross 
section (dotted line) without form factors.}}
\label{fig5}
\end{figure}

\section{Conclusions}

We have studied the cross section of $J/\psi$ dissociation 
by $K^*$ in a meson-exchange model that
includes pseudo-scalar-pseudo-scalar-vector-meson couplings,
three-vector-meson couplings, pseudo-scalar-vector-vector-meson couplings
and four-point couplings. Off-shell effects at the vertices were handled 
with QCD sum rule estimates for the form factors. The inclusion of anomalous 
parity interactions (pseudo-scalar-vector-vector-meson couplings) has opened
additional channels to the absorption mechanism. Their contribution
are very important especially for large values of the initial energy,
$\sqrt{s}>4.7~\GeV$.

As shown in Fig.~2 our results, without form factors, have the same energy 
dependence of $J/\psi$ absorption by $\rho$ from ref.~\cite{osl}. The
inclusion of the form factors changes the energy dependence of the
absorption cross section in a nontrivial way, as shown in Fig.~3. This 
modification in the energy dependence is
similar to what was found in ref.~\cite{haga2} for $J/\psi$ absorption by 
$\rho$.

With QCD sum rules estimates for the coupling constants and form factors,
the total $J/\psi-K^*$ cross section was found to 
be $1.6 \sim1.9$ mb for
$4.2\leq\sqrt{s}\leq5~\GeV$. Using the same form factors and QCD sum rules
to estimate the
value for $g_{\rho DD}$ we get for the $J/\psi - \rho$ total absorption
cross section $3.0 \sim10.0$ mb, in the same energy range.

\section*{Acknowledgments}

This work was supported by CNPq and FAPESP.

\end{document}